\documentclass[lettersize,journal]{IEEEtran}
\usepackage{amsmath,amsfonts}
\usepackage{algpseudocode} 
\usepackage{algorithm}
\usepackage{array}
\usepackage[caption=false,font=normalsize,labelfont=sf,textfont=sf]{subfig}
\usepackage{textcomp}
\usepackage{stfloats}
\usepackage{url}
\usepackage{verbatim}
\usepackage{graphicx}
\usepackage{cite}
\usepackage[inkscapearea=page]{svg}
\usepackage{adjustbox}
\usepackage{enumitem}
\usepackage{titlesec}

\newcommand{\GC}{\emph{Green Cores}}
\newcommand{\OGC}{\emph{openstack-gc}}
\newcommand{\RealTimeClouds}{\emph{Real-Time Clouds}}
\newcommand{\RDCO}{\emph{Renewables-driven cores}}
\newcommand{\MacroVMExecutionModel}{VM Execution Model}
\newcommand{\MacroVMPackingAlgorithm}{VM Packing Algorithm}

\newcommand{\AlgoEvctCountOursReductionOverBestfit}{$79.64\%$}

\newcommand{\AlgoRnwHarvestOursIncreaseOverCritaware}{$34.83\%$}

\titlespacing*{\section}{0pt}{0.4\baselineskip}{0.4\baselineskip}
\titlespacing*{\subsection}{0pt}{0.4\baselineskip}{0.4\baselineskip}

\begin{document}

\title{A Framework for Carbon-aware Real-Time Workload Management in Clouds using Renewables-driven Cores}

\author{Tharindu B. Hewage, Shashikant Ilager, Maria A. Rodriguez, and Rajkumar Buyya
    \thanks{T. B. Hewage, M. A. Rodriguez, R. Buyya are with the Cloud Computing and Distributed Systems (CLOUDS) Laboratory, School of Computing and Information Systems, University of Melbourne, Parkville, VIC 3010, Australia.}%
    \thanks{S. Ilager is with the Informatics Institute, University of Amsterdam.}
}

\maketitle

\begin{abstract}
Cloud platforms commonly exploit workload temporal flexibility to reduce their carbon emissions. They suspend/resume workload execution for when and where the energy is greenest. However, increasingly prevalent delay-intolerant real-time workloads challenge this approach. To this end, we present a framework to harvest green renewable energy for real-time workloads in cloud systems. We use \RDCO\ in servers to dynamically switch CPU cores between real-time and low-power profiles, matching renewable energy availability. We then develop a \MacroVMExecutionModel\ to guarantee running VMs are allocated with cores in the \emph{real-time power profile}. If such cores are insufficient, we conduct criticality-aware VM evictions as needed. Furthermore, we develop a \MacroVMPackingAlgorithm\ to utilize available cores across the data center. We introduce the \GC\ concept in our algorithm to convert renewable energy usage into a server inventory attribute. Based on this, we jointly optimize for renewable energy utilization and reduction of VM eviction incidents. We implement a prototype of our framework in OpenStack as \OGC. Using an experimental \OGC\ cloud and a large-scale simulation testbed, we expose our framework to VMs running RTEval, a real-time evaluation program, and a 14-day Azure VM arrival trace. Our results show: i) a $6.52\times$ reduction in coefficient of variation of real-time latency over an existing workload temporal flexibility-based solution, and ii) a joint \AlgoEvctCountOursReductionOverBestfit\ reduction in eviction incidents with a \AlgoRnwHarvestOursIncreaseOverCritaware\ increase in energy harvest over the state-of-the-art packing algorithms. We open source \OGC\ at \url{https://github.com/tharindu-b-hewage/openstack-gc}.
\end{abstract}


\vspace{-2pt}

\section{Introduction}

\par Data centers consumed approximately 1-1.3\% of global electricity demand in 2022 \cite{iea23dc-and-networks}. Between 2015 and 2022, data center energy usage increased by 20-70\% \cite{iea23dc-and-networks}. The recent unprecedented compute demand due to Artificial Intelligence (AI) and Machine Learning (ML) workloads indicates that this trend will continue to grow \cite{bianchini2024dc-power-past-present-future}. Data centers often connect to electricity grids with shares of energy generation based on fossil fuels. As a result, in 2020, data centers were responsible for 0.9\% of energy-related greenhouse gas (GHG) emissions \cite{iea23dc-and-networks}. Climate crisis-driven road maps necessitate that data center emissions drop by half by 2030 to meet global Net Zero Emissions goals \cite{iea23dc-and-networks, bianchini2024dc-power-past-present-future}.

\par In response to GHG emissions, electrical grids continue to integrate low-emission renewable energy sources. In 2022, the share of renewables in total electricity generation was 39\% and is projected to be 91\% by 2035 \cite{iea23low-emmision-sources}. However, growing variable-availability (intermittent) renewable energy sources, such as solar and wind, challenge electrical grids \cite{bianchini2024dc-power-past-present-future}. Between 2022 and 2035, energy reports project the share of solar and wind renewables in total generation to rise from 12\% to 58\% \cite{iea23low-emmision-sources}.

\par Data centers develop various \textit{load matching} strategies to match workload execution over intermittent renewable energy. Amongst them, \textit{load shifting} is commonly practised \cite{sukprasert2024loadshiftingcloud, masanet2020recalibratingdatacenterenergyestimates, radovanovic23datacentercarbonaware, zheng2020mitigating}. \textit{Load shifting} uses workloads with temporal flexibility to suspend/resume their execution. For example, Google's delay-tolerant workloads, such as machine learning, data compaction, and data processing, tolerate delays as long as their work gets completed within 24 hours \cite{radovanovic23datacentercarbonaware}. Workloads execute in periods when renewable energy capacity is higher, resulting in reduced GHG emissions. However, \textit{load shifting} falls short when applied to real-time workloads with strict response time boundaries \cite{barbieri2023what-is-rt}. Real-time workloads cannot tolerate the delays inherent in \textit{load shifting}.

\par Nevertheless, the growing prevalence of real-time cloud applications, such as autonomous vehicles, industrial automation \cite{zhang_adaptive_2019}, and railway control systems \cite{gala2021rt-cloud} expects to account for nearly 30\% of the world data by 2025 \cite{idc2018digitizationofworld}. As a result, cloud operators will eventually have to incorporate growing real-time workloads in intermittent renewable energy integration. In this context, one must find an alternative \emph{load matching} strategy to \textit{load shifting} for delay-intolerant real-time workloads. Existing solutions, such as applying CPU-wide \emph{low power profile} to match renewable energy supply \cite{piga2024dvfs-boost}, often result in increased latency, making them unsuitable for real-time applications. Moreover, techniques like Harvest Virtual Machines (HVMs), which allow uninterrupted execution of workloads with reduced resources \cite{agarwal2023slackshed}, can still degrade performance and fail to meet real-time constraints.

\par Given these challenges, there is a need for an efficient strategy to integrate renewable energy into real-time cloud systems (\RealTimeClouds). To this end, we propose a framework to harvest renewable energy in \RealTimeClouds. We use \RDCO\ to integrate renewable energy at the server level. It dynamically switches the power profiles of each CPU core between a \emph{real-time power profile} and a \emph{low power profile} to match renewable energy intermittency. Then, our framework applies a twofold solution to utilize this dynamic core availability. First, we develop a \textbf{\MacroVMExecutionModel} to guarantee that real-time virtual machines (VMs) occupy cores at the \emph{real-time power profile}. Our model adopts renewable energy fluctuations by conducting criticality-aware VM evictions as needed. Secondly, we develop a \textbf{\MacroVMPackingAlgorithm} to optimize the use of available cores across the data center. It reduces the likelihood of VM evictions while maximizing renewable energy utilization (renewable energy harvest). Our algorithm frames renewable energy management as a VM placement optimization problem by introducing the concept of \textbf{\GC}. \GC\ presents each server as an inventory of two virtual CPU core types: Green and Regular. Green cores quantify renewable energy usage, whereas Regular cores quantify core usage that does not increase risks of VM eviction incidents. Using \GC, we achieve a computationally inexpensive VM packing algorithm, which is required to handle VM throughput at the data center level \cite{radovanovic23datacentercarbonaware}.

\par We implement our framework in OpenStack \cite{openstack} as \OGC. We combine OpenStack’s control plane with an on-node daemon service. The daemon service implements \RDCO\ in the server using per-core sleep states. \OGC’s control plane then communicates with the daemon service to orchestrate our \MacroVMExecutionModel\ and \MacroVMPackingAlgorithm. We evaluate our framework at the server level using VMs running RTEval, a program from the Real-Time Linux project to measure real-time performance \cite{rteval}. We evaluate our framework at the data center level using a 14-day VM arrival trace from Azure \cite{azure2020packingtrace}. We use two testbeds: an experimental \OGC\ cloud deployed on an HPE ProLiant server with a 12-core Intel Xeon CPU, and a large-scale simulation testbed. We make the following contributions in designing, implementing, and evaluating our framework.

 \begin{itemize}[leftmargin=*]
    \item We propose a server level \textbf{\MacroVMExecutionModel} to utilize renewable energy without degrading real-time latency performance in VMs. We leverage criticality-aware VM evictions for that.
    \item We propose a data center level \textbf{\MacroVMPackingAlgorithm} to reduce VM eviction incidents over renewable energy utilization.
    \item We implement a prototype of our framework in OpenStack, detailing its design and demonstrating its practicality.
    \item We evaluate our approach against multiple baselines. Our results show: i) $6.52\times$ reduction in coefficient of variation of real-time latency in VMs over the existing workload temporal flexibility-based VM execution model, and ii) a joint optimization of \AlgoEvctCountOursReductionOverBestfit\ reduction in VM eviction incidents and \AlgoRnwHarvestOursIncreaseOverCritaware\ increase in utilized renewable energy over state-of-the-art packing algorithms \cite{kumhare2021poweroversubscription}.
\end{itemize}

The rest of the paper is organized as follows: Section \ref{sec::bk} provides the background and motivation for our problem with a use case study. Section \ref{sec::dn} details our system model and problem formulation. Section \ref{sec::design} outlines the design of our proposed framework. Section \ref{sec::impl} describes the implementation of \OGC. Section \ref{sec::pe} presents the performance evaluation of our framework. Section \ref{sec::rw} discusses related work, and Section \ref{sec::cn} concludes the paper and outlines future work.

\section{Background, Motivation and Use case}
\label{sec::bk}

This section provides background on real-time workloads in clouds (\RealTimeClouds) and the application of \RDCO. It then motivates our contributions with a use case study. Finally, it outlines the key takeaways.

\subsection{Real-Time Clouds}

The prevalence of real-time use cases is estimated to account for nearly 30\% of the world’s data in 2025  \cite{idc2018digitizationofworld}. In clouds, real-time systems are emerging across use cases such as industry 4.0, autonomous vehicles, and software-defined networks \cite{ubuntu2020ctoguideforrealtimekernel}. Clouds provide the flexibility of virtualized resource management and cost-effective scaling over dedicated hardware-based solutions \cite{gala2021rt-cloud}. In a real-time cloud system (\RealTimeClouds), components are deployed using VMs \cite{gala2021rt-cloud,zhang_adaptive_2019,casini2021rt-hypervisor-latency}. Within a VM, the correctness of a computation is measured based not only on the logical correctness of the result but also on the time at which it was produced \cite{ubuntu2020ctoguideforrealtimekernel}. \RealTimeClouds\ tune the entire virtualization stack to reduce latency in application execution \cite{intel2023realtimereadiness,openstack2023realtime,intel2016nfv-tunning,gala2021rt-cloud}, such as disabling CPU power optimization features to achieve a consistent power profile in CPU cores, pinning each VM core to a server core to avoid CPU time sharing latency, and using real-time-ready kernels to execute real-time application calls without delays \cite{ubuntu2020ctoguideforrealtimekernel}. Moreover, \RealTimeClouds\ deploy an application-specific middleware layer for system management, such as providing fault tolerance in the events of VM failures \cite{gala2021rt-cloud,etsiosm2020autoheal}.

\subsection{Renewables-driven Cores}

\RDCO\ is a load-matching technique that dynamically changes the power profile in CPU cores. \RDCO\ provide the opportunity to avoid CPU-wide power capping/throttling in matching server load for intermittent renewable energy utilization \cite{li2011solar-core}. When the server load exceeds the energy supply, \RDCO\ sets cores into a \emph{low power profile} as needed to reduce the server load below the energy supply. Therefore, cores not in the \emph{low power profile} can change with time, requiring a workload execution model capable of utilizing this. For example, existing works that apply \RDCO\ for VM workloads use Harvest Virtual Machine (HVM) to dynamically change the number of physical cores available to a VM \cite{ambati2020harvestvms}.

\subsection{ Motivation}
\label{sec::mt}

We outline the motivation behind our proposed framework, specifically focusing on the rationale for selecting \RDCO\ as the load-matching technique for \RealTimeClouds. Then, we discuss the lack of static compute allocation in the existing VM execution solution for \RDCO\ and how it can impact real-time VMs. Following this, we present our approach to addressing this limitation by exploiting the presence of mixed-criticality in \RealTimeClouds, composed of VMs hosting \emph{critical} components and \emph{best-effort} components, to implement criticality-aware VM evictions.

\noindent\textbf{How does Renewables-driven cores align with the needs of Real-Time Clouds?} Renewable energy integration at the server level using \RDCO\ as the load-matching technique enables avoiding both suspend/resume and CPU-wide throttling of workloads. Figure \ref{fig:mt-rdco} illustrates a scenario where a set of cores reside in the \emph{low power profile}. However, the remaining cores reside in the \emph{real-time power profile}. They provide the opportunity to serve real-time VMs amidst renewable energy fluctuations.

\begin{figure}[t]
    \centering
    \includesvg[width=0.9\columnwidth]{figures/mt-rdco.svg}
    \caption{\RDCO\ in \RealTimeClouds}
    \label{fig:mt-rdco}
\end{figure}

\noindent\textbf{Why is a static compute allocation important to real-time VMs?:} To apply \RDCO\ in \RealTimeClouds, we need a workload execution model to match dynamic core availability for VMs. In this regard, the existing solution is Harvest VMs (HVMs) \cite{agarwal2023slackshed}. HVM’s approach is to change the number of physical cores (pCPUs) in the VM but preserve the number of virtual cores (vCPUs). It allows continued execution of the VM amidst dynamic core availability. However, for real-time VMs, this dynamic compute allocation can introduce performance degradation. To observe its impact, we evaluate the real-time performance of an HVM with combinations of different vCPUs and pCPUs. Our experimental HVM consists of 2 vCPUs. We set pCPUs to the \emph{real-time power profile}. We then execute the RTEval \cite{rteval} program inside the HVM to measure the real-time performance. 

\par
\begin{figure}[b]
    \centering
    \includesvg[width=0.7\columnwidth]{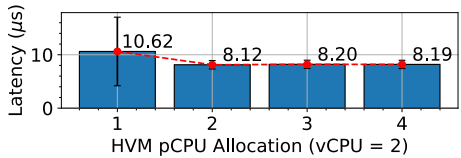}
    \caption{Comparison of real-time latency performance of a two-core Harvest VM (HVM) \cite{agarwal2023slackshed} over different physical CPU core allocations.}
    \label{fig:mt-rt-perf-with-affinity}
\end{figure}

 The results are illustrated in Figure \ref{fig:mt-rt-perf-with-affinity}. When the number of pCPUs $\geq$ to the number of vCPUs, the HVM sustains a consistent real-time latency performance, having both mean and mean absolute deviation statistics consistent for all three cases of pCPUs $\geq 2$. The opposite shows increased latency variance inside the HVM. Compared to pCPUs $= 2$, the case of pCPUs $= 1$ increases the mean latency by $30\%$ alongside a $7.32\times$ increase of the mean absolute deviation of latency. Therefore, insufficient pCPU allocations can lead to undesirable real-time latency spikes in the VMs. Maintaining a static compute allocation is important to avoid such scenarios.

\noindent\textbf{Opportunities in mixed-criticality within Real-Time Clouds to provide static compute allocation for VMs over Renewables-driven cores:} An alternative to the HVM approach of continuing the execution of VMs with insufficient pCPUs is to evict the VMs. Existing works show opportunities for this in \RealTimeClouds\ via the mixed-criticality of real-time systems \cite{durrieu2016dreams,gala2021rt-cloud,etsiosm2020autoheal}. Firstly, they model real-time system components as either \emph{critical} or \emph{best-effort}. Then, an application-specific middleware layer exploits this mixed-criticality to provide fault tolerance for component failures. In real-time cloud systems, application-specific middleware layers use reconfiguration policies to recover the system upon VM failures \cite{etsiosm2020autoheal}. In this context, there is an opportunity to conduct VM evictions in \RealTimeClouds\ safely. Since a VM eviction is a well-defined failure event, the application-specific middleware layer can tolerate it through reconfiguration. More importantly, we can guarantee a static compute allocation for VMs with a fixed allocation of CPU cores and conduct criticality-aware VM eviction if cores are insufficient instead of continued VM execution with degraded performance.

\begin{table}[t]
    \caption{Mixed-criticality use cases in 5G Network Slicing \cite{zhang_adaptive_2019}}
    \label{tab:mt-slicing}
    \centering
    \begin{tabular}{|c|c|c|} \hline 
         \textbf{Scenario}&  \textbf{Reliability}& \textbf{Criticality}\\ \hline 
         Autonomous Driving&  99.999\%& Critical\\ \hline 
         Industrial Machinery&  99.999\%& Critical\\ \hline 
         4K/8K HD Video&  -& Best-effort\\ \hline 
         Mass Gathering&  -& Best-effort\\ \hline
    \end{tabular}
\end{table}

\subsection{Usecase}
\par
\begin{figure}[b]
    \centering
    \includesvg[width=0.7\columnwidth]{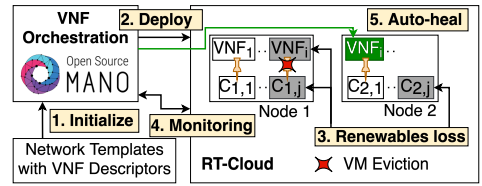}
    \caption{Use case: understanding the effect of load matching with evictions via application-level reconfiguration. 5G network slicing prototype of open source NFV Management and Orchestration (MANO) with OpenStack as the virtualized infrastructure provider (i.e. real-time cloud provider). MANO’s auto-healing feature facilitates application-level reconfiguration over VM failures \cite{etsiosm2020autoheal}.}
    \label{fig:mt-mano}
\end{figure}

 To further motivate our approach, we experiment with a real-time cloud use case of 5G Network Slicing via Virtual Network Functions (VNFs) \cite{zhang_adaptive_2019}. As the application-specific middleware layer, we employ the production-grade VNF management and orchestration middleware, OSM MANO \cite{etsiosm2020autoheal}. We map VNFs to \emph{critical} and \emph{best-effort} components based on the service quality level of their network slice. Table \ref{tab:mt-slicing} denotes an example where the criticality of four different 5G scenarios is interpreted based on service reliability. Figure \ref{fig:mt-mano} illustrates our study. We connect the MANO deployment with a real-time tuned two-node OpenStack deployment as the real-time cloud. We use the auto-heal feature of MANO as the reconfiguration policy \cite{etsiosm2020autoheal}. Each VNF is deployed as a VM in OpenStack with two virtual CPU cores. We use 50\% \RDCO\ in servers at 100\% initial renewable energy capacity. Once the deployment stabilizes, we drop that to 0\%, reducing server core count by half and evicting VMs to load match. We repeat the experiment for two VM scheduling approaches in OpenStack.

\par Table \ref{tab:mt_autoheal} denotes our observations. Firstly, upon the loss of renewable energy capacity, MANO reconfigured the available cores through auto-healing. Secondly, OpenStack initiates VM evictions. Thus, knowledge of VM criticality is beneficial in reducing the impact of eviction. For example, it permits evicting \emph{best-effort} components prior to \emph{critical} components. Thirdly, the VM packing strategy can change the number of VM evictions. Of the two packing approaches used in our use case, the Tightly approach triggered two eviction incidents, whereas the spreading approach yielded none. 

\begin{table}[t]
    \centering
    \caption{Comparison of the VM packing inventory over different packing strategies as 5G network slicing prototype conduct load matching for renewable energy loss via VM evictions.}
    \label{tab:mt_autoheal}
    \begin{tabular}{|m{1.1cm}|m{0.85cm}|m{0.85cm}|m{0.85cm}|m{0.85cm}|m{1.7cm}|} 
        \hline 
        \textbf{Packing} &  
        \multicolumn{2}{|c|}{\textbf{Before}} &
        \multicolumn{2}{|c|}{\textbf{After}}  &
        \textbf{Evictions} \\ 
        \hline 
         &  \textbf{Node 1}&  \textbf{Node 2}& \textbf{Node 1}&  \textbf{Node 2}  & \\ 
         \hline 
         Tightly
&  $0/8$ &  $8/8$ & $4/4$ &  $4/4$ & 2\\ \hline 
         Spread
&  $4/8$ &  $4/8$ & $4/4$ &  $4/4$  & 0\\ \hline
    \end{tabular}
\end{table}

\subsection{Key Takeaways}

From our motivations and the use case experiment, we identify the following key takeaways:

\begin{enumerate}[leftmargin=*]
    \item \label{enm:takeaways:point-1} \RDCO\ enable integrating renewable energy in \RealTimeClouds. In that,
    \begin{enumerate}[leftmargin=*]
        \item A static compute allocation needs to be maintained for VMs to avoid real-time latency spikes.
        \item Criticality-aware VM evictions provide an opportunity to maintain a static compute allocation for VMs.
    \end{enumerate}
    \item \label{enm::takeaways:point-2} The data center VM packing strategy can influence the likelihood of VM eviction incidents.
\end{enumerate}

Motivated by the above, we design our framework to apply \RDCO\ in \RealTimeClouds. Our approach addresses the point \ref{enm:takeaways:point-1} using a \MacroVMExecutionModel, and the point \ref{enm::takeaways:point-2} using a \MacroVMPackingAlgorithm.

\section{System Model and Problem Formulation}
\label{sec::dn}

\subsection{System Model}

\begin{figure}[t]
    \centering
    \includesvg[width=\linewidth]{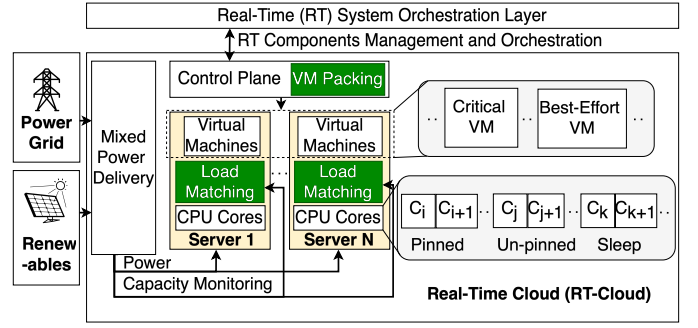}
    \caption{A high-level system model of the proposed carbon-aware real-time cloud. We highlight components with our contributions in green.}
    \label{fig:dn_sys-model}
\end{figure}

\noindent Our system model is shown in Figure \ref{fig:dn_sys-model}. We employ server-level renewable energy integration, each server receiving dedicated allocations of grid and renewable energy capacities through a mixed power delivery system. Allocations are even across all servers. We use homogeneous servers for simplicity, but the model can be adapted for heterogeneous servers by allocating power capacities proportionately. Each server monitors the dynamic availability of its allocated renewable capacity for \textit{load matching}. We model renewable energy as intermittent and carbon-free and grid energy as static and carbon-intensive. An application-specific middleware layer manages each VM. It can tolerate VM eviction incidents. At arrival, VMs provide their criticality to the cloud control plane as either \emph{critical} or \emph{best-effort}. A VM packing algorithm then places VMs in the data center server inventory.

\subsection{Problem Formulation}
\noindent\textbf{Server power modelling:} Groundwork from prior studies state server power is CPU dominant and can be estimated over 90\% accuracy using a linear function ($f$) of CPU power ($P_{CPU}(t)$) \cite{radovanovic2022dcpowermodel}. Thus, we define server power at time $t$ ($P_{S}(t)$) as,

\[P_{S}(t) = f(P_{CPU}(t))\]

In multi-core CPUs, the cumulative sum of core power becomes a close upper bound of $P_{CPU}(t)$ \cite{basmadjian2012multicorepowermodel}. Based on this, we derive an upper-bound to server power and use that as an estimate to $P_{CPU}(t)$. Therefore, for a server with $N$ number of cores,

\begin{equation}
    \label{eq::spm:sp-with-cores}
    P_{S}(t) \simeq f(\sum_{i=1}^{N} P_{CORE_i}(t))
\end{equation}

where $P_{CORE_i}(t)$ is the power consumption of $i^{th}$ core at time $t$. Commodity servers often consist of homogeneous cores. Thus, we apply the same model here. Next, we model power states for the three distinct states of $P_{CORE_i}(t)$. When a core is unused, its power state is either,

\begin{itemize}[leftmargin=*]
    \item \textit{Active} $\equiv P_{CORE_i}(t) = P_{ACT}$ (an idle core)
    \item \textit{Sleep} $\equiv P_{CORE_i}(t) = P_{SLP}$ (a core in the \emph{low power profile})
\end{itemize}

 In contrast, a core pinned to a VM exhibits a power state of $P_{CORE_i}(t) = F(U_{CORE_i}(t))$. Where $U_{CORE_i}(t)$ is the utilization of the core at time $t$, and $F$ is a linear function \cite{basmadjian2012multicorepowermodel}. Dynamics of $U_{CORE_i}(t)$ depends on the VM workload, which is a black box to the cloud operator \cite{kumhare2021poweroversubscription}. Therefore, in packing problems, a representative utilization statistic is commonly estimated based on historical data \cite{kumhare2021poweroversubscription,radovanovic2022dcpowermodel}. Based on this, we use $U_{RT}$ to estimate $U_{CORE_i}(t)$. Cloud operator sets the exact $U_{RT}$ value using deployment-specific data. As a result, the pinned power state of a core becomes,

\par\noindent\textit{Pinned} $\equiv P_{CORE_i}(t) = P_{PIN}$, where $P_{PIN} = F(U_{RT})$

\par
\begin{figure}[b]
    \centering
    \includesvg[width=1\linewidth]{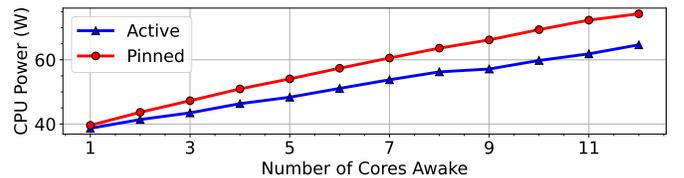}
    \caption{CPU power as cores awake in \RDCO: measured with an Intel Xeon Silver CPU where core power states are: i) \textit{Sleep} $\equiv$ sleep state of C6, ii) \textit{Active} $\equiv$ sleep state of $POLL$, and iii) \textit{Pinned} $\equiv$ pinned with 100\% utilization}
    \label{fig:dn_core-pw-states}.
\end{figure}

 We verify our core power model with an Intel Xeon Silver CPU with 12 cores. For that, we use $U_{RT} = 100\%$. We wake up cores from 1 to 12 and plot CPU package power obtained through Intel’s RAPL interface. We conduct the same experiment for both \textit{Active} and \textit{Pinned} scenarios. Figure \ref{fig:dn_core-pw-states} illustrates our results. Graphs that sustain linear trends with constant slopes, corresponding to $P_{PIN}$ and $P_{ACT}$. The same trends imply the $P_{SLP}$. 

\par We then apply the core power model in Equation \ref{eq::spm:sp-with-cores} and derive the following model to estimate server power using core counts as variables.

\begin{equation}
    \label{eq:pm-cores}
    \begin{split}
        P_{S}(t) \simeq f(m(t) \times P_{PIN} + l(t) \times P_{SLP} \\
        + (N - m(t) - l(t)) \times P_{ACT})
    \end{split}
\end{equation}
where at time $t$, $m(t)$ is the pinned core count and $l(t)$ is the sleeping core count. 

\noindent\textbf{Renewable energy harvest:} In our model, servers are allocated with dedicated grid and renewable energy capacities. When server power meets grid capacity, we denote $P_{S}(t) = P_{GRID}$. Renewable energy harvesting begins when $P_{S}(t) > P_{GRID}$. Therefore, for an arbitrary time period $\Delta T$, we denote harvested renewable energy ($E_{RW}(\Delta T)$), 

\begin{equation}
    \label{eq::rw-harvest}
    \begin{split}
        E_{RW}(\Delta T) = \int_{\Delta T}^{} \{u(P_{S}(t) - P_{GRID}) \\ \times (P_{S}(t) - P_{GRID})\} \, dt
    \end{split}
\end{equation}
where $u$ is the unit step function.

\noindent \textbf{Service quality:} We model service quality with real-time latency performance of VMs and the number of VM eviction incidents. We prefer a lower value in both.

\noindent \textbf{Problem formulation:} We formulate our problem as follows: Given an arbitrary $\Delta T$ period, maximize renewable energy harvesting while preserving service quality. Thus our \textbf{objective} is: 

    \[
    \text{Maximize} \quad E_{RW}(\Delta T) \quad \text{and} \quad \text{Minimize} \quad n
    \]
    where $E_{RW}(\Delta T)$ is the harvested renewable energy derived in Equation \ref{eq::rw-harvest}, and $n$ is the number of VM eviction incidents. The objective function should satisfy the following \textbf{constraints}: 
    
    \[
    \bar{l}_i \leq \bar{l}_{\text{max}_i} \quad \text{and} \quad \sigma_i \leq \sigma_{\text{max}_i} \quad \text{for} \quad vm_i \in S_{vm}
    \]
    $S_{vm}$ is the virtual machines executed during $\Delta T$ and $\bar{l}_i$ and $\sigma_i$ are the mean and variance of real-time latency, respectively. $\bar{l}_{\text{max}_i}$ and $\sigma_{\text{max}_i}$ are deployment-specific upper bounds.

\section{Design}
\label{sec::design}

In this section, we outline the design of our framework. It combines a server level \MacroVMExecutionModel\ with a data center level \MacroVMPackingAlgorithm.

\subsection{Design of the Server level \MacroVMExecutionModel}

We select a subset of cores at the server level and apply \RDCO\ to them. At 100\% renewable energy capacity, we set all server cores to the \emph{real-time power profile}. At 0\% renewable energy capacity, we put all cores in the subset to a \emph{low power profile}. For in-between, we set the \emph{real-time power profile} to a partial amount of cores in the subset and set the \emph{low power profile} for the rest. In this case, the number of cores in the \emph{real-time power profile} is proportionate to available renewable energy capacity. For example, at 50\% renewable energy capacity, half of the cores in the subset are set to the \emph{real-time power profile}. In our approach, \RDCO\ dynamics depend solely on the renewable energy intermittency and are independent of the workload execution dynamics.

\par We pin VM cores to server cores set to the \emph{real-time power profile} at the VM deployment and do not change it for the duration of the VM lifetime. If the number of such server cores is insufficient to serve running VMs, we perform a minimum amount of criticality-aware VM evictions. We evict \emph{best-effort} VMs first and \emph{critical} VMs as a last resort. Our model guarantees a static compute allocation for a VM’s lifetime. The VM eviction events trigger well-defined VM failure events at the application-specific middleware layer, allowing it to recover through reconfiguration. In the next section, we design a data center-level \MacroVMPackingAlgorithm\ to reduce the possibility of such server-level VM eviction events.

\subsection{Design of the Data Center level \MacroVMPackingAlgorithm}

Possibilities of VM eviction incidents with our \MacroVMExecutionModel\ increases when the number of VMs packed in a server begin renewable energy harvesting (see Equation \ref{eq::rw-harvest}). Therefore, optimizing VM eviction incidents must be conducted jointly with optimizing the renewable energy harvest. We convert the joint optimization task into a VM packing optimization problem. Firstly, we introduce \GC: a concept to convert server-level renewable energy usage into a packing attribute. Then, using \GC, we derive a data center-level VM packing algorithm.

\noindent\textbf{Server inventory of Green Cores:} We denote the number of cores that remain in a constant \emph{real-time power profile} as $R$ where $0 \leq R \leq N$, such that the size of \RDCO\ at time $t$ ($l(t)$) is $0 \leq l(t) \leq N - R$. We choose a value for $R$ such that when $R$ cores are at a \textit{Pinned} power state and $l(t)$ is $N-R$, the server power draw meets the grid capacity. Using our server power model in Equation \ref{eq::spm:sp-with-cores} we derive,
\begin{equation}
    \label{eq:grid-capacity}
    P_{S}(t) \simeq f(R \times P_{PIN} + (N - R) \times P_{SLP}) = P_{GRID}
\end{equation}
where $m(t) = R$, $l(t) = N - R$, and $P_{GRID}$ is the grid capacity.

We derive an equation for the amount of renewable energy harvested by subtracting Equation \ref{eq:grid-capacity} from Equation \ref{eq::spm:sp-with-cores}.

\begin{equation}
\label{eq:int:rnw-draw}
    \begin{split}
        P_{S}(t) - P_{GRID} = f(m(t) \times P_{PIN} + l(t) \times P_{SLP} \\
        + (N - m(t) - l(t)) \times P_{ACT} \\ - R \times P_{PIN} - (N - R) \times P_{SLP})
    \end{split}
\end{equation}

We then model $m(t)$ with $R$ as \( m(t) = R + g(t) \) where \( g(t) \) is an arbitrary function. Substituting this model in Equation \ref{eq:int:rnw-draw} yields:

\begin{equation*}
    \begin{split}
        P_{S}(t) - P_{GRID} = f(g(t) \times (P_{PIN} - P_{ACT}) \\
        + (P_{ACT} - P_{SLP}) \times ((N - R) - l(t)))
    \end{split}
\end{equation*}

Here, we denote the leakage power ($L(t)$): power drawn by \RDCO\ at the \textit{Active} state as \( L(t) = (P_{ACT} - P_{SLP}) \times ((N - R) - l(t)) \).

\begin{equation}
\label{eq:gc-1}
    \begin{split}
        P_{S}(t) - P_{GRID} = f(g(t) \times (P_{PIN} - P_{ACT}) + L(t))
    \end{split}
\end{equation}

Then, substituting Equation \ref{eq:gc-1} in Equation \ref{eq::rw-harvest}, we estimate renewable energy harvest for an arbitrary time period $\delta t$ where $g(t) \geq 0$,

\[E_{RW}(\delta t) = \int_{\delta t}^{} f(g(t) \times (P_{PIN} - P_{ACT}) + L(t)) \, dt\]
where $f$ is the linear function to map CPU power into server power, in which a positive input yields a positive power value. Here, when $\delta t$ is small enough to match the measurement interval of the system, the $E_{RW}(\delta t)$ can be stated as,

\[E_{RW}(\delta t) \simeq f(g(\delta t) \times (P_{PIN} - P_{ACT}) + L(\delta t)) \times \delta t \]
where both $g(\delta t)$ and $L(\delta t)$ are values measured for the $\delta t$ interval. The design of our \RDCO\ ensures that $L(\delta t)$ is independent from workload execution. In contrast, the value of $g(\delta t)$ depends on the packing decisions of the cloud’s control plane. Therefore, if $g(\delta t)$ is changed with different packing algorithms,

\begin{equation}
\label{eq:rnw-change-with-packing}
    E_{RW}(\delta t) \propto g(\delta t)
\end{equation}

Equation \ref{eq:rnw-change-with-packing} states that if the packing algorithm positively increases $g(\delta t)$, the renewable energy harvest increases. However, a positive $g(\delta t)$ also means that $m(t) > R$, implying VMs are pinned to \RDCO, thus increasing the eviction possibilities. We then use the $g(\delta t)$ derivations to derive a server inventory called \GC. \GC\ presents a server with CPU cores of two kinds: Green and Regular. For Green cores, active cores ($C_{G_{active}}$) are calculated with $(N - R) - l(t)$ and used cores ($C_{G_{used}}$) are calculated with $g(t)$ if $g(t) \geq 0$ (otherwise is set to $0$). For Regular cores, active ($C_{R_{active}}$), and used ($C_{R_{used}}$) cores are calculated with $R$, and $m(t)$ if $m(t) < R$ (otherwise is set to $R$), respectively. Calculations of Green cores quantify the usage of renewable energy capacity, and calculations of Regular cores quantify the usage of cores that do not increase the risks of VM eviction incidents.

\begin{figure}[t!]
    \centering
    \includesvg[width=\columnwidth]{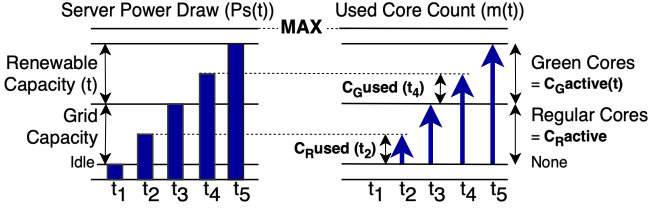}
    \caption{Comparison of server power draw vs proposed server inventory of \GC.}
    \label{fig:virt-pack-inv}
\end{figure}

\begin{figure}[b]
    \centering
    \includesvg[width=0.46\columnwidth]{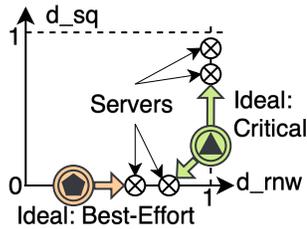}
    \caption{Joint optimization of renewable energy harvest and VM eviction incidents via the distance to ideal points in the Euclidean space of the proposed server inventory of \GC.}
    \label{fig:servers-2d}
\end{figure}

\par This is illustrated in Figure \ref{fig:virt-pack-inv} with a side-by-side comparison between power domain and the server inventory of \GC. In this scenario, the number of pinned cores ($m(t)$) increases from $t_1$ to $t_5$. As a result, server power draw ($P_{S}(t)$) increases. Until $t_3$, server power draw is less than the grid capacity, where $C_{G_{used}} = 0$ and $C_{R_{used}}$ is proportionate to the utilized energy capacity. During this period, there are no risks of VM eviction incidents. Beyond $t_3$, the risk of VM eviction incidents increases as the server power draw utilizes renewable energy capacity, where $C_{G_{used}}$ is proportionate to the utilized energy capacity and $C_{R_{used}}$ is capped at $R$.

\noindent\textbf{\MacroVMPackingAlgorithm:} Using the server inventory of \GC, we define a 2-dimensional feature vector to represent a server $\equiv (d_{rnw},d_{sq})$. We denote $d_{rnw}$ to signal the opportunity to increase renewable energy harvest and calculate it as \(d_{rnw}(t) = \frac{|C_{G_{active}(t)} - C_{G_{used}}(t)|}{C_{G_{active}}(t)}\), and we denote $d_{sq}(t)$ to signal the opportunity to reduce VM eviction incidents and calculate it as \(d_{rnw}(t) = \frac{|C_{G_{active}(t)} - C_{G_{used}}(t)|}{C_{G_{active}}(t)}\).

\par Then, as illustrated in Figure \ref{fig:servers-2d}, we plot each data center server in the 2-dimensional Euclidean space of the feature vector. Additionally, we mark two reference points in this space, called \emph{ideal points},  to represent an ideal server for each \emph{critical} and \emph{best-effort} VM type. When a VM creation request arrives, we choose the corresponding ideal point, sort data center servers based on the distance to that ideal point, and select the closest one as the packing decision. Algorithm \ref{algo:pk} outlines the pseudo-code for this. Ideal points make our algorithm tunable, enabling it to cater to deployment-specific performances \cite{ubuntu2020ctoguideforrealtimekernel}. Our performance evaluation section demonstrates that using a real VM arrival trace.

\begin{algorithm}[t!]
	\caption{Proposed \MacroVMPackingAlgorithm}\label{algo:pk}
	\begin{algorithmic}[1]
        \Function{GetPlacementPreferences}{$V$: VM, $S$: Candidate servers, $\tau_1$: Ideal point for \emph{critical} VMs, $\tau_2$: Ideal point for \emph{best-effort} VMs}
            \State $\epsilon \gets GetCriticality(V)$
            \State $\tau \gets GetIdealPoint(\epsilon, \tau_1, \tau_2)$
            \ForAll{$s_{i} \in S$}
                \State $d_{sq_i} \gets GetRNW(s_{i})$
                \State $d_{rnw_i} \gets GetSQ(s_{i})$
                \State $d_{i} \gets GetDistance(d_{sq_i},d_{rnw_i}, \tau)$
                \State $s_{i}.score \gets 1 - d_{i}$
            \EndFor
            \State \Return $getSorted(S)$
        \EndFunction
        \Function{GetIdealPoint}{$\epsilon, \tau_1, \tau_2$}
            \State \Return $\tau_1$ if $\epsilon$ is \emph{critical} else $\tau_2$
        \EndFunction
        \Function{GetSQ}{$s_{i}$}
            \State $C_{R_{active}}(t), C_{R_{used}}(t) \gets s_{i}$
            \State \Return $\frac{|C_{R_{active}}(t) - C_{R_{used}}(t)|}{C_{R_{active}}(t)}$
        \EndFunction
        \Function{GetRNW}{$s_{i}$}
            \State $C_{G_{active}}(t), C_{G_{used}}(t) \gets s_{i}$
            \State \Return $\frac{|C_{G_{active}(t)} - C_{G_{used}}(t)|}{C_{G_{active}}(t)}$
        \EndFunction
        \Function{GetDistance}{$d_{sq_i},d_{rnw_i}, \tau$}
            \State $d_{sq_\tau}, d_{rnw_\tau} \gets \tau$
            \State $distance \gets \frac{\sqrt{(d_{sq_\tau} - d_{sq_i})^2 +(d_{rnw_\tau} - d_{rnw_i})^2}}{\sqrt{2}}$
            \State \Return $distance$
        \EndFunction
	\end{algorithmic}
    \label{pk}
\end{algorithm}

\section{Implementation}
\label{sec::impl}

\begin{figure}[t]
    \centering
    \includesvg[width=0.86\columnwidth]{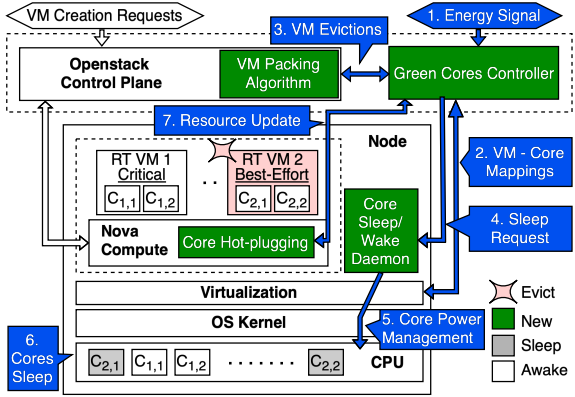}
    \caption{System architecture of OpenStack-GC. It outlines the server load matching workflow for intermittent availability of renewable energy.}
    \label{fig:openstack-gc-architecture}
\end{figure}

In this section, we outline \OGC: implementation of our framework in OpenStack.

 \par\noindent\textbf{Implementation of Openstack-GC:}  
 Figure \ref{fig:openstack-gc-architecture} illustrates the system architecture of \OGC. We highlight newly added OpenStack extensions in green. We deploy an on-node daemon service to realize \RDCO. We introduce a Green Cores Controller at the control plane to orchestrate the proposed \MacroVMExecutionModel. We implement the proposed \MacroVMPackingAlgorithm\ as a VM scheduling algorithm in OpenStack.
 
\par\noindent\textbf{Renewables-driven cores:} We implement an on-node daemon service in Golang to control per-core power profiles. By making an API call to the daemon service, the \OGC\ control plane can specify the number of cores to put into a specific power profile. If the \emph{real-time power profile} is requested, the daemon service sets the requested number of cores into a high-performance state. If the  \emph{low power profile} is requested, the daemon service sets the requested cores into a deep sleep state. The daemon service wraps the Intel Power Optimization Library\footnote{\url{https://github.com/intel/power-optimization-library.git}} and overrides the kernel management of the sleep state and operating frequency of each core to achieve this.

\par\noindent\textbf{\MacroVMExecutionModel:} We orchestrate our \MacroVMExecutionModel\ using the load matching workflows of \OGC. First, we enable the dedicated cores feature in OpenStack to pin each VM core to a dedicated server core, resulting in a static core allocation for each deployed VM. Then, our load-matching workflows of \OGC\ take place. Suppose an increased energy capacity signal arrives to \OGC. In that case, the Green Cores Controller calculates and notifies on-node daemon services to set the required number of cores from the \emph{low power profile} to the \emph{real-time power profile}. If a decreased energy capacity signal is provided to \OGC, the Green Cores Controller pings APIs of the virtualization layer (\OGC\ uses Libvirt\footnote{\url{https://www.libvirt.org}}) in each node to obtain mappings of VM cores to server cores. Then, the Green Cores Controller calculates and triggers criticality-aware VM evictions by blocking API calls to the OpenStack. Upon completion, the required cores are put to the \emph{low power profile} using on-node daemon services. Figure \ref{fig:openstack-gc-architecture} illustrates the workflow for the decreased energy capacity. In both cases, our modified Nova Compute, OpenStack’s on-node compute service, periodically polls the Green Cores Controller to obtain cores at the \emph{low power profile}. Afterwards, Nova Compute signals the control place to omit cores from VM scheduling in the \emph{low power profile}.

\par\noindent\textbf{\MacroVMPackingAlgorithm:} We modify the OpenStack scheduler service to poll the Green Cores Controller and obtain server inventory attributes of \GC\ for all server nodes. To provide that, the Green Cores Controller pings virtualization layers of servers to obtain core usage information and calculates server inventory attributes of \GC. Our implementation of the proposed \MacroVMPackingAlgorithm\ as a VM scheduling algorithm in OpenStack consumes obtained \GC\ server attributes to make VM placement decisions.

\section{Performance Evaluation}
\label{sec::pe}

\begin{table}[t]
    \caption{Openstack-GC Prototype: Node Specifications}
    \label{tb:ogc-node}
    \centering
        \begin{tabular}{|m{0.48\columnwidth}|m{0.40\columnwidth}|} \hline 
            \textbf{Attribute} & \textbf{Description}\\ \hline 
            Server Model& ProLiant DL380 Gen10\\ \hline 
            CPU &Intel(R) Xeon(R) Silver 4214\\ \hline 
            Physical Cores&12\\ \hline 
            Hyper Threading&Disabled\\ \hline 
            Renewables-driven Cores &6\\ \hline
            Real-time power profile&C-state = $POLL$ at 2699 MHz\\ \hline 
            Low power profile &C-state = C6\\ \hline
        \end{tabular}
\end{table}

\begin{table}
\centering
\caption{Openstack-GC Prototype: VM Specifications}
\label{tab:rt-vm-specs}
    \begin{tabular}{|m{0.28\columnwidth}|m{0.6\columnwidth}|} 
        \hline 
        \textbf{Attribute} & \textbf{Description} \\ \hline 
 Resources&CPU: 6 Cores, RAM: 6GB\\
        \hline 
        OS & CentOS 7\\ 
        \hline 
        Kernel& Linux 3.10.0 + CERN’s Real-Time patches\\ 
        \hline 
        System Load & Load test of RTEval \cite{rteval}\\
        \hline
        Latency Monitoring& Cyclictest \cite{cyclictest}\\ \hline
    \end{tabular}
\end{table}

To evaluate our server level \MacroVMExecutionModel, we use a prototype \OGC\ deployment. We use trace-driven simulations at scale to evaluate our data center level \MacroVMPackingAlgorithm.

\subsection{Experimental Setup}

\par\noindent\textbf{Server level: OpenstackGC prototype experiments}

We deploy a prototype \OGC\ cloud on an HPE ProLiant server with a 12-core Intel Xeon Silver CPU. Table \ref{tb:ogc-node} outlines the node's specifications.

\par\noindent\textbf{Workload}: Table \ref{tab:rt-vm-specs} outlines the VM specifications for our real-time workloads. We use CentOS 7 VMs with CERN’s real-time kernel patches \cite{cernrtlinux} applied. We run the RTEval tool from the Linux foundation project, Real-Time Linux \cite{rteval}, to emulate a system load. Alongside the load, RTEval continuously measures the VM kernel’s real-time performance via the Cyclictest tool \cite{cyclictest}.

\par\noindent\textbf{Baseline}:  We use Harvest Virtual Machines (HVM): the existing VM execution model over \RDCO\ \cite{agarwal2023slackshed} to evaluate advancements of our \MacroVMExecutionModel.

\par\noindent\textbf{Metrics}: We use Intel’s Running Average Power Limit \cite{intel2017sdmanual} interface to capture CPU metrics in the server every 0.5 seconds. We collect i) CPU package power consumption, ii) core residencies at the C6 sleep state, and iii) core operating frequency in MHz. Inside VMs, we measure the latency to wake up a real-time thread using the Cyclictest tool \cite{cyclictest}.

\par\noindent\textbf{Data center level: trace-driven simulations at scale}

\begin{figure*}[htbp]
    \centering
    \includesvg[width=0.8\linewidth]{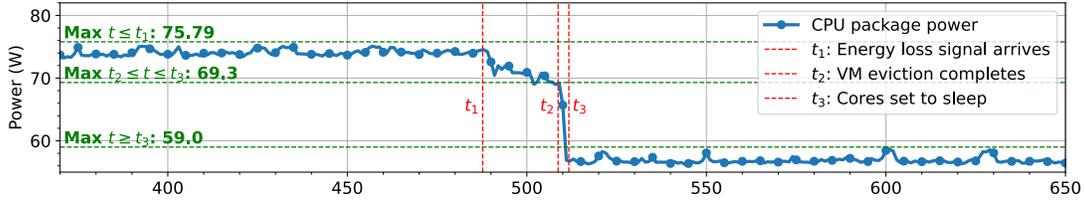}
     \caption{CPU package power measurements obtained through Intel’s RAPL interface \cite{intel2017sdmanual} during the load matching experiment of the OpenStack-GC prototype. The proposed VM Execution Model manages the server load over the renewable energy availability. Before $t_1$, the server executes two 6-core VMs. An energy loss signal at $t_1$ triggers the eviction of one VM to unpin six cores, completing at $t_2$. The unpinned cores enter into deep sleep at $t_3$.}
    \label{fig:rslt:load-matching}
\end{figure*}

We use 8K+ servers, each with 40 CPU cores, to match the realistic similar values in Microsoft's Azure's cloud zones \cite{hadary2020protean}. Existing fallback mechanisms of Azure's data center power delivery suggest that a 12\% power overdraw is manageable \cite{kumhare2021poweroversubscription}. To operate within that, we add four cores in each server and use them as \RDCO. We use solar energy dynamics from the ELIA dataset \cite{eliaopendata}. We normalize and scale the trace so that the maximum renewable energy capacity can wake all \RDCO\ in a server.

\par\noindent\textbf{Workload:} We use VM workload data on Microsoft's Azure Compute from Azure's 14-day packing trace \cite{azure2020packingtrace}. It contains dynamics of VM request arrivals, VM resource requirements, VM lifetime on Azure, and VM evictability. We categorize evictable VMs as \emph{best-effort} VMs and the rest as \emph{critical} VMs. 

\par\noindent\textbf{Baselines}: To evaluate the proposed VM packing algorithm, we use two comparison baselines. The \emph{Best-Fit packing (best-fit)} is a commonly used packing approach in production clouds \cite{hadary2020protean,openstack-scheduling} that packs VMs tightly in servers. We use it to evaluate our advancements over a commonly used data center packing approach. The \emph{Criticality-Aware packing (crt-aware)} is a packing approach that reduces VM throttling incidents in power over-subscribed data centers \cite{kumhare2021poweroversubscription}. Similar to our problem context, it leverages VM criticality to reduce VM performance impact incidents incurred from server load exceeding available power capacity. We use it to evaluate our advancements over the state-of-the-art.

\par\noindent\textbf{Metrics}: We use the \emph{Harvested Renewables} to measure utilization of renewable energy capacity. Using derivations of \GC, for a period of $T$, we calculate it as $\equiv\int_0^T C_{G_{used}}(t) dt$. We use the \emph{Eviction Incidents} to count the number of eviction incidents of \emph{best-effort} and \emph{critical} VMs. We use the \emph{Normalized Lifetime (nLT)} to measure the severity of an eviction incident. For each evicted VM, we normalize its lifetime from the original lifetime in the trace. A lower nLT value implies increased severity.

\subsection{Evaluation of Server-level \MacroVMExecutionModel}

We evaluate the proposed VM Execution Model's ability to maintain the server load to match available renewable energy capacity and its impact on the real-time performance of VMs. Firstly, we signal \OGC\ prototype deployment with a 100\% energy capacity to wake all \RDCO\ in the server. Then, we pin all cores by deploying two 6-core VMs. In both VMs, we run the RTEval program for the duration of the experiment to emulate a peak load. Then, we signal a 0\% energy capacity.

\par Figure \ref{fig:rslt:load-matching} shows the CPU package power observed throughout. We collect it via Intel's Running Average Power Limit (RAPL) \cite{intel2017sdmanual} interface. The CPU package draws up to 75.79W at peak load with a relatively constant trend. $t_1$ denotes the arrival of the energy loss signal for 0\% renewable energy capacity. \OGC's response shows a two-stage power reduction; $t_1$ - $t_2$ and $t_2$ - $t_3$. The former shows the power reduction from evicting one of the VMs to unpin six cores. Latter shows the power reduction from putting unpinned cores to deep sleep. After $t_3$, CPU package power does not exceed 59W. It translates to a 22\% reduction of the peak power draw. With the linear model of CPU power to server power \cite{radovanovic23datacentercarbonaware}, our \OGC\ deployment shows a reduction of 22\% of the server peak power in matching 100\% to 0\% loss of renewable energy capacity.

\par
\begin{figure}[b!]
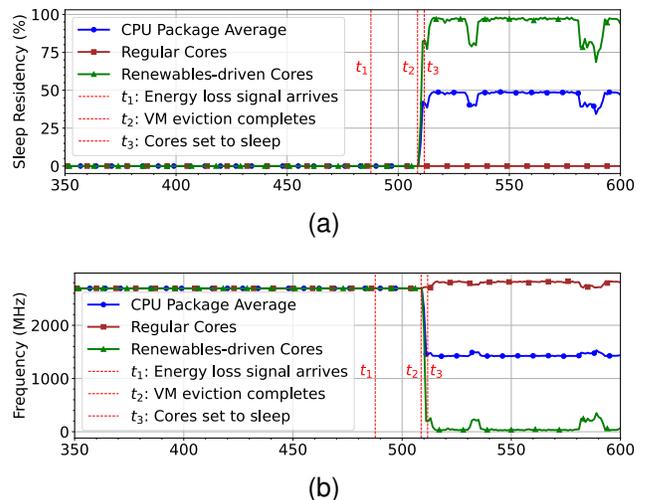

    \centering
    \subfloat[]{\includesvg[width=0.95\linewidth]{results/sleep-residency.svg}} \\
    \subfloat[]{\includesvg[width=0.95\linewidth]{results/op-frq.svg}}
    \caption{CPU core power characteristics in Openstack-GC prototype during server load matching for renewable energy loss. We obtain power metrics through Intel’s RAPL interface \cite{intel2017sdmanual}. (a) CPU core C6 deep sleep state residency. (b) CPU core operating frequency.}
    \label{fig:rslt:load-matching:cpu-metrics}
\end{figure}

 Figure \ref{fig:rslt:load-matching:cpu-metrics} shows CPU core power characteristics. We capture operating frequency and C6 deep sleep state residency (i.e. in a given measurement period, the percentage that the core resided in the sleep state) of cores. We average it for the six cores that enter deep sleep state after $t_3$ (Renewables-driven cores), for the six cores that continue to operate, and for overall. Until $t_2$, \OGC\ maintains a constant 2700 MHz operating frequency of cores with 0\% residency in deep sleep, showing cores operating at the \emph{real-time power profile}. Afterwards, Renewable-driven cores mostly reside in a deep sleep with 0 MHz operating frequency, showing their \emph{low power profile}. It shows that \OGC\ only changes power profiles after the VM eviction completion at $t_2$. Throughout the lifetimes of VMs, VM cores are allocated with physical cores in the \emph{real-time power profile}. 

\par The spikes in maintaining the \emph{low power profile} show the characteristics of controlling core power through Intel's Power Optimization library that we use in \OGC. Our prototype also has an overhead of running external services, including the OpenStack control plane services in the same server, which could be attributed to the spikes shown. Despite that, the server power draw shows a 59W upper bound in figure \ref{fig:rslt:load-matching}, showing \OGC\ can maintain a constant power reduction over the intermittent spikes in the \emph{low power profile}.

\par Next, we evaluate the impact of the proposed VM Execution Model at the application layer for the real-time workloads. We use the Harvest VM (HVM) as a comparison baseline, the existing VM execution model over \RDCO \cite{agarwal2023slackshed}. We use an experimental deployment of OSM MANO \cite{etsiosm2020autoheal}, a Virtual Network Functions (VNF) orchestration and management application layer, to match a real-time application layer having both \emph{critical} and \emph{best-effort} components (see Table \ref{tab:mt-slicing}). In public clouds, server utilization is around 60\%, and the packing density (i.e. utilization of servers running at least one VM) is around 85\% \cite{hadary2020protean}. To match that, we use two 12-core servers and tightly pack one server with two 6-core VMs while the other is left unused. MANO is a generic orchestration layer where the exact time-bound requirements depend on the use case. Therefore, for VMs, we measure the real-time latency of the VM kernel for a consistent real-time latency performance independent of the use case. To match application-level reconfiguration over component failures, we enable MANO's auto-heal feature, which reconfigures itself via VM redeployment. HVM executes VMs under resource variations. To match it's worst-case, we set the dynamics of Renewables-driven cores to sleep five cores in both servers, such that HVM executes a VM with 6 VM cores allocated to one physical core. In contrast, the proposed approach evicts one of the VMs, triggering MANO to redeploy it in the unused server. We obtain the time taken for reconfiguration via MANO's event logs.

\par
\begin{figure}[htbp]
    \centering
    \includesvg[width=\linewidth]{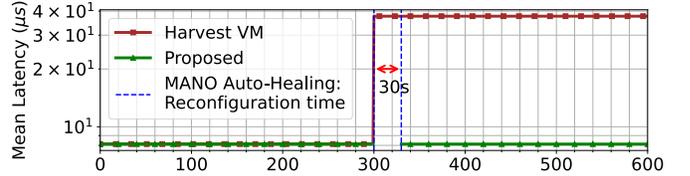}
    \caption{Application layer real-time performance comparison of VM execution models. Experimental setup deploys VNF orchestration application, OSM MANO \cite{etsiosm2020autoheal}, over two 12-core server nodes of \OGC\ prototype. We plot the mean real-time latency obtained through RTEval \cite{rteval} running in the affected VMs. MANO's auto-heal feature is enabled to reconfigure itself upon VM eviction by redeploying. Initially, two 6-core VMs tightly pack a server where the other is left unused to match resource usages of public clouds \cite{hadary2020protean}. At $t = 300$, worst-case resource reduction with an HVM is emulated by sleeping five cores in each server. The proposed model evicts a VM, where HVM reduces the physical core count to one in a VM. }
    \label{fig:hvm-shrink}
\end{figure}

 Figure \ref{fig:hvm-shrink} shows the real-time latency performance of the affected VM in both the proposed and HVM approaches. In both methods, the remaining VM continues executing under the same core allocations without any performance impact since the server's physical core count is sufficient. Until the server core sleep event at time axis $=$ 300, the affected VM in both approaches shows the same mean real-time latency. Afterwards, the core count reduces. With HVM, the affected VM's mean real-time latency increases from 8.13$\mu s$ to 37.65$\mu s$. When comparing the coefficient of variation of the VM's real-time latency, it increases to 6.52$\times$. With the proposed approach, the affected VM undergoes a 30-second service unavailability. However, when it resumes afterwards, the VM retains the same real-time latency performance. The results show that, unlike the existing temporal flexibility-based approach, the proposed VM Execution Model maintains intact real-time latency performance. In doing so, it incurs brief service unavailability from VM evictions as a trade-off. In the next section, we evaluate the role of our proposed VM packing algorithm in reducing the impact of that on the application layer.

\subsection{Evaluation of Data center Level \MacroVMPackingAlgorithm}\label{sec:perf_packingalgo}

\begin{figure*}[htbp]
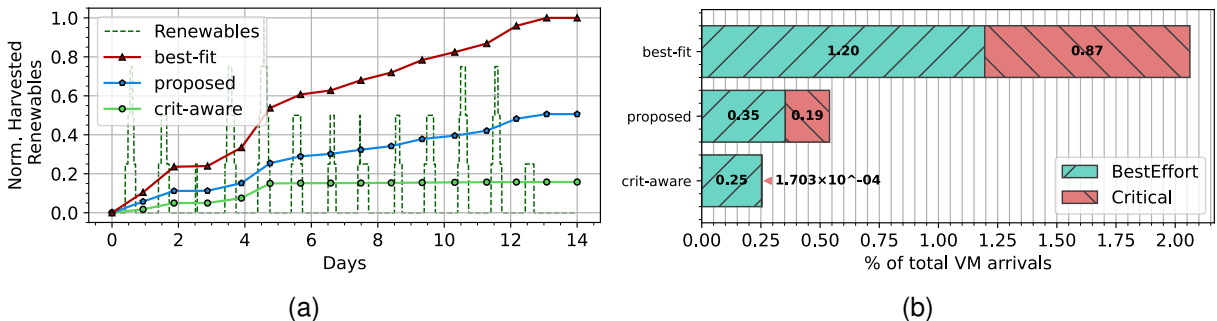

    \centering
    \subfloat[]{\includesvg[width=0.45\linewidth]{results/sim/full/full_harvest.svg}}
    \subfloat[]{\includesvg[width=0.45\linewidth]{results/sim/full/full_evictions.svg}}
    \caption{Performance of the proposed \MacroVMPackingAlgorithm\ in packing a 14-day Azure VM arrival trace \cite{azure2020packingtrace} to jointly optimize renewable energy harvest and VM eviction incidents. (a) Accumulation of harvested renewable energy capacity. (b) The number of VM eviction incidents.}
    \label{fig:rslt:full}
\end{figure*}

We use a large-scale simulation test bed to evaluate our framework at the data center scale. In the test bed, we first implement \RDCO\ with a trace of renewable energy dynamics and then implement the proposed VM Execution Model for server-level load matching. We expose the test bed to a 14-day Azure VM workload arrival trace. We use our proposed VM packing algorithm and comparison baselines to determine VM allocations across servers.

\par We first tune our algorithm by observing its performance over short-running experiments. We aim to jointly optimize the reduction of VM eviction incidents and increase renewable energy harvest. Once tuned, we conduct 14-day packing experiments. Figure \ref{fig:rslt:full} shows (a) harvested renewable energy and (b) the number of VM evictions. Values for the former are normalized among the comparison baselines, and values for the latter are expressed as a percentage of the total number of VM requests. 

\begin{figure*}[htbp]
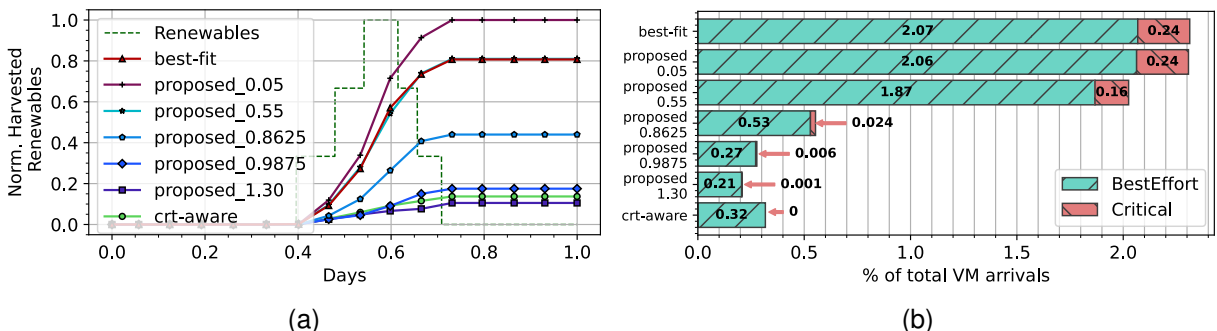

    \centering
    \subfloat[]{\includesvg[width=0.45\linewidth]{results/sim/sa/harvest.svg}}
    \subfloat[]{\includesvg[width=0.45\linewidth]{results/sim/sa/evictions.svg}}
    \caption{Performance of the proposed packing algorithm during its sensitivity analysis. We change the distance between its ideal point parameters and conduct a packing experiment of 24 hours in each step. (a) Accumulation of harvested renewable energy capacity. (b) The number of VM eviction incidents.}
    \label{fig:rslt:sa}
\end{figure*}

\par
\begin{figure}[htbp]
    \centering
    \includesvg[width=0.9\linewidth]{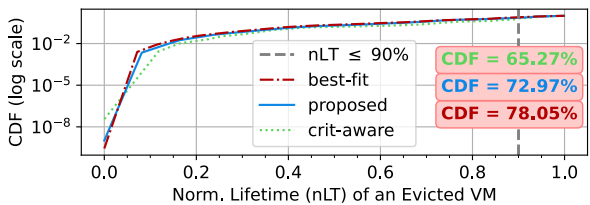}
    \caption{Distribution of normalized lifetime (nLT) of evicted VMs during the 14-day VM packing experiment.}
    \label{fig:rslt:full:nlt}
\end{figure}

 Both baselines show their inability to conduct joint optimization. The best-fit algorithm is most effective in harnessing renewable energy yet evicts over 2\% of VM requests. It incurs the highest amount of \emph{critical} VM evictions among the three algorithms. The crit-aware, on the other hand, shows the most effectiveness in reducing VM eviction incidents with 0.25\% of total VMs evicted with a 1.703$\times10^{-4}\%$ of \emph{critical} VM evictions, the lowest amongst three algorithms. However, it shows the least harvested renewable energy with an 80\% reduction from the best-fit algorithm. Our proposed algorithm shows a joint optimization, a \AlgoRnwHarvestOursIncreaseOverCritaware\ increase over crit-aware in harvested renewable energy and a \AlgoEvctCountOursReductionOverBestfit\ reduction of VM eviction incidents compared to best-fit. Our algorithm reaches 50\% of the renewable energy harvest performance of best-fit with a 26.09\% VM eviction incidents of best-fit, showing its joint optimization characteristic to favour lesser eviction incidents. Figure \ref{fig:rslt:full:nlt} shows distributions of a normalized lifetime (nLT) of evicted VMs. The proposed VM packing algorithm surpasses best-fit and approaches crit-aware with the CDF value for nLT $\leq 90\%$.

\par\noindent\textbf{Sensitivity analysis:} We conduct a sensitivity analysis of our algorithm’s hyper-parameters. In the proposed packing algorithm, we represent each server using a 2-dimensional feature vector $\equiv (d_{rnw}, d_{sq})$: $d_{rnw}$ quantifies renewable energy usage and $d_{sq}$ quantifies the possibility of VM eviction incidents. Parameters of our algorithm are two instances of this vector (called ideal points), one for each \emph{critical} and \emph{best-effort} VM type. For initial values, we set \emph{critical} VM ideal point to $(1,0.5)$ such that those VMs prefer servers with the potential to reduce VM eviction incidents, and \emph{best-effort} ideal point to $(0.2,0.0)$ such that those VMs prefer servers with the potential to harvest renewable energy. 

\par In subsequent experiments, we move \emph{critical} ideal point closer to the other and conduct 24-hour packing experiments in each step. Figure \ref{fig:rslt:sa} illustrates our results. As ideal points move closer, our proposed algorithm favours increasing renewable energy harvest, surpassing the leading baseline best-fit at a distance of $0.05$. Although this behaviour compromises eviction incidents, the number of incidents is still less than that of the best-fit. In contrast, as ideal points deviate, our proposed algorithm favours decreasing eviction incidents, surpassing the leading baseline crt-aware at distances of $0.9875$ and $1.30$.

\subsection{Discussion}

\par Our evaluations show the potential of our framework to manage server power using \RDCO. Per-core application of \emph{low power profile} demonstrates our framework can reduce the server power to match supply variations of renewable energy. CPU power metrics shown during that indicate that if a core pins to a VM, that core's power profile transition will not occur. Even if unused cores are insufficient, the framework evicts the VM first before changing the power profile. As a result, our framework guarantees a static compute allocation throughout a VM's lifetime. Evaluation of the real-time latency performance of VM kernels shows that the static compute allocation provided in our framework significantly outperforms existing workload temporal-flexibility-based VM execution solutions. 

\par Our approach shows two trade-offs. Firstly, a sustained core power profile until the completion of VM evictions requires redundancies in the data center power delivery to support the short periods of server power overdraws. However, existing cloud data centers can support similar requirements \cite{kumhare2021poweroversubscription}. Therefore, our framework fits into existing data center designs. Secondly, the static compute allocation requires VM evictions if enough unused cores are unavailable to match the energy supply. However, an application-specific middleware layer in clouds manages real-time VMs, which provides fault tolerance over VM evictions \cite{etsiosm2020autoheal,gala2021rt-cloud}. Therefore, VM evictions in our framework do not incur application-level failures for real-time workloads. Moreover, large-scale packing experiments show that our framework can reduce the number of VM eviction incidents by utilizing core availability across the data center servers, jointly optimizing that with the utilization of renewable energy capacity. Our eviction-based approach exploits findings of a previous study showing that cloud applications prefer VM evictions over continued VM execution with performance degradation \cite{kumhare2021poweroversubscription}.

\par Performance of our framework improves with the presence of \emph{best-effort} VMs. Therefore, cloud operators need to tune our algorithm according to the workload variations. Sensitivity analysis of our framework's packing algorithm parameters shows the ability to support that (see Section \ref{sec:perf_packingalgo}). The algorithm can be tuned to favour renewable energy harvesting for a deployment that expects an increased number of \emph{best-effort} VMs. Otherwise, it can be tuned down to reduce the number of VM eviction incidents. Our framework design expects server utilization levels in typical data centers, where a slack of unused capacity is available \cite{hadary2020protean}. It allows the real-time application layer to reconfigure in the events of VM evictions. If the data center utilization levels are much higher, the application layer may be unable to do so. In such cases, the algorithm tuning must be adjusted to reduce eviction incidents.

\begin{table*}[htbp]
    \centering
        \caption{Comparison of Relevant Work with Ours}
        \label{tab::lit-compare}
        \begin{tabular}{ m{0.18\textwidth}  m{0.13\textwidth}   m{0.09\textwidth} m{0.10\textwidth}m{0.09\textwidth}m{0.12\textwidth}m{0.10\textwidth}}
        \hline
        \textbf{Work} & \centering\arraybackslash\textbf{Renewables-driven \newline Cores} & \centering\arraybackslash\textbf{Critical \newline Workloads} & \centering\arraybackslash\textbf{Real-Time \newline Readiness} & \centering\arraybackslash\textbf{VM Execution} & \centering\arraybackslash\textbf{Criticality-aware \newline Packing} & \centering\arraybackslash\textbf{Renewables Harvest}\\
        \hline
        SolarCore (2011) \cite{li2011solar-core} & \centering\arraybackslash\checkmark &  & & & & \centering\arraybackslash\checkmark \\ 
        \hline
        Chameleon (2013) \cite{li2013chameleon} & \centering\arraybackslash\checkmark & & & & & \centering\arraybackslash\checkmark \\ 
        \hline
        Kumbhare et. al (2021) \cite{kumhare2021poweroversubscription} & & \centering\arraybackslash\checkmark & & \centering\arraybackslash\checkmark & \centering\arraybackslash\checkmark & \\ 
        \hline
        PowerMorph (2022) \cite{jahanshahi2022powermorph} & \centering\arraybackslash\checkmark & \centering\arraybackslash\checkmark & & & & \centering\arraybackslash\checkmark \\ 
        \hline
        Slackshed (2023) \cite{agarwal2023slackshed} & \centering\arraybackslash\checkmark & & & \centering\arraybackslash\checkmark & & \centering\arraybackslash\checkmark \\ 
        \hline
        Our Proposed & \centering\arraybackslash\checkmark & \centering\arraybackslash\checkmark & \centering\arraybackslash\checkmark & \centering\arraybackslash\checkmark & \centering\arraybackslash\checkmark & \centering\arraybackslash\checkmark \\ 
        \hline
        \end{tabular}
\end{table*}

\section{Related Work}
\label{sec::rw}

\par \noindent \textbf{Load matching with renewables-driven cores:} Common load matching techniques for intermittent renewable energy, such as geographical load balancing, workload migration, admission control, and capacity planning \cite{radovanovic23datacentercarbonaware,masanet2020recalibratingdatacenterenergyestimates,zheng2020mitigating}, depend on either suspending/resuming or migrating flexible workloads. In contrast, \RDCO\ avoids both by performing load matching with dynamic core availability. SolarCore \cite{li2011solar-core} and Chameleon \cite{li2013chameleon} use per-core Dynamic Voltage and Frequency Scaling (DVFS) and Power Gating to implement \RDCO. In their work, workloads utilizing power-adjusted cores can undergo performance degradation, thus better suited for throughput workloads with flexible deadlines. PowerMorph \cite{jahanshahi2022powermorph} improves this via core grouping, hosting \emph{critical} and \emph{best-effort} workloads and power adjustments isolated to core groups. However, workload core affinity can dynamically change during load matching, unfavourable for time-critical workloads such as real-time compute \cite{openstack2023realtime}. Slackshed \cite{agarwal2023slackshed} implement \RDCO\ for virtual machine (VM) execution. They achieve uninterrupted VM execution at the expense of dynamic CPU allocation, thus better suited for throughput workloads with flexible time constraints. In contrast, our work preserves workload time boundaries over \RDCO\ and leverages criticality-aware VM evictions within safe limits of the application layer.

\par \noindent \textbf{VM packing algorithms:} VM packing is a widely studied research problem. Most existing works focus on variants of bin packing algorithms to improve resource utilization at scale \cite{hadary2020protean, baker2018cloud}, yet consider servers as static inventories. Opposed to that, Kumbhare et al. \cite{kumhare2021poweroversubscription} explore an inventory where servers oversubscribe power delivery, yielding a dynamically changing inventory capacity. They propose a criticality-aware packing algorithm to co-pack \emph{critical} and \emph{best-effort} components, reducing workload impact. However, in doing so, they do not consider renewable energy harvesting opportunities. In contrast, our work achieves joint optimization of workload impact and harvesting in dynamic inventories.

\section{Conclusions and Future Work}
\label{sec::cn}

\par In this work, we proposed a framework to harvest renewable energy with real-time workloads in clouds using \RDCO. Existing works address only flexible workloads and fail to preserve the time bounds of real-time computing. To address that, we used a two-fold design of: i) \MacroVMExecutionModel\ for server-level load matching via criticality-aware VM evictions, and ii) \MacroVMPackingAlgorithm\ for data center level packing to reduce VM eviction incidents. We implemented our framework in OpenStack as \OGC. We deployed a prototype \OGC\ cloud to evaluate at the server level and used trace-driven simulations to evaluate at the data center level. As evidenced through empirical results, the static compute allocation provided by our framework demonstrated its superiority in real-time workloads by reducing the coefficient of variation of real-time latency in VMs by $6.52\times$ over the existing workload temporal-flexibility-based solution. Furthermore, our proposed framework showcased its safe energy harvesting capability with a joint \AlgoEvctCountOursReductionOverBestfit\ reduction of VM eviction incidents and \AlgoRnwHarvestOursIncreaseOverCritaware\ increase of harvested renewable energy over state-of-the-art baselines. Moreover, our sensitivity analysis of parameters demonstrated its capacity to cater to specific harvesting requirements.

\par In the future, we intend to integrate prediction-based utilization modelling of VM workloads to advance the server power model in our framework. Additionally, since power profile transitions in our framework can impact CPU temperature, it is worth exploring the thermal impact of our framework on server longevity to help reduce embodied carbon in the data center.

\par\noindent\textbf{Software availability:}  \OGC\ has been open-sourced at \url{https://github.com/tharindu-b-hewage/openstack-gc}.

\bibliographystyle{IEEEtran}
\bibliography{references}

\vspace{-33pt}
\begin{IEEEbiography}[{
\includegraphics[width=1in,height=1.25in,clip,keepaspectratio]{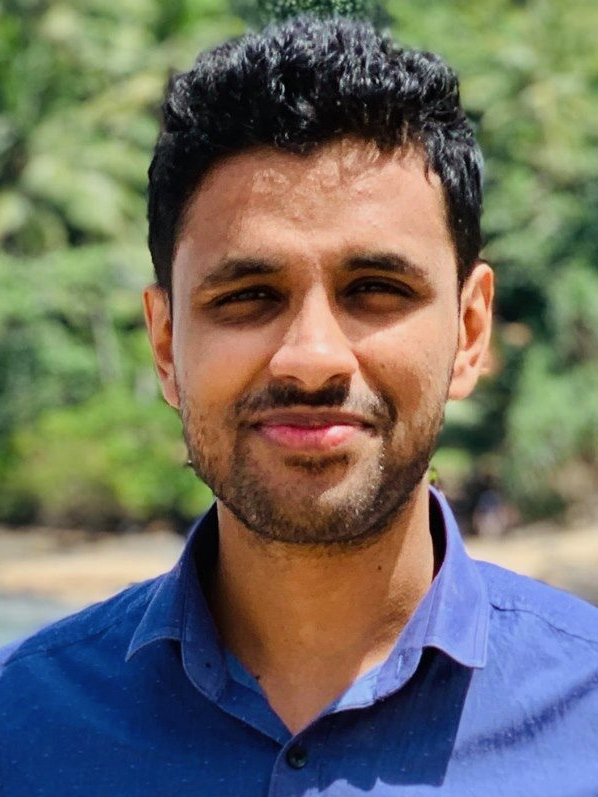}
}]
{Tharindu B. Hewage} is working toward a PhD with the Cloud Computing and Distributed Systems (CLOUDS) Laboratory, Department of Computing and Information Systems, University of Melbourne, Australia. His research interests include distributed systems and cloud computing. His current research focuses on energy and carbon-aware resource management in edge-cloud systems.
\end{IEEEbiography}

\vspace{-30pt}
\begin{IEEEbiography}[{
\includegraphics[width=1in,height=1.25in,clip,keepaspectratio]{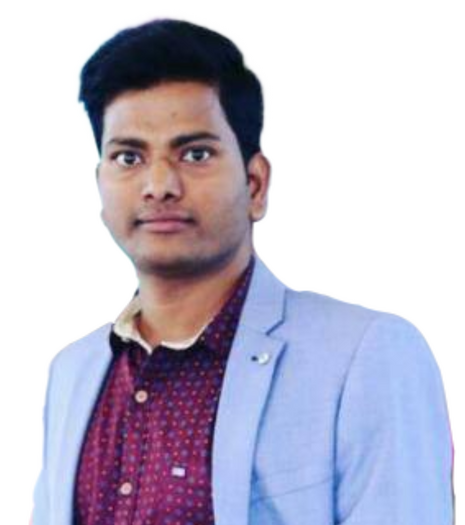}
}]
{Shashikant Ilager} is an assistant professor at the Informatics Institute, University of Amsterdam, Netherlands. He is a member Multiscale Networked Systems research group. He works at the intersection of distributed systems, energy efficiency, and machine learning. His recent research explores the energy efficiency and performance optimization of data-intensive and distributed AI applications.
\end{IEEEbiography}

\vspace{-30pt}
\begin{IEEEbiography}[{
\includegraphics[width=1in,height=1.25in,clip,keepaspectratio]{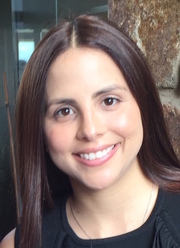}
}]
{Maria Rodriguez Read} is a lecturer in the School of Computing and Information Systems, University of Melbourne, Australia. Her research interests lie in the field of distributed and parallel systems. Her recent research works involve investigating how containerized and cloud-native applications can be better supported by cloud providers to offer users advantages in terms of reduced cost and more scalable, robust, and flexible application deployment.
\end{IEEEbiography}

\vspace{-30pt}
\begin{IEEEbiography}[{
\includegraphics[width=1in,height=1.25in,clip,keepaspectratio]{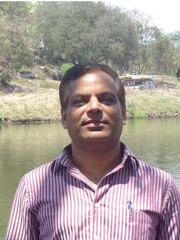}
}]
{Rajkumar Buyya} (Fellow, IEEE) is a Redmond Barry distinguished professor and director of the Cloud Computing and Distributed Systems (CLOUDS) Laboratory, University of Melbourne, Australia. He has authored over 800 publications and seven textbooks. He is one of the highly cited authors in computer science and software engineering worldwide (h-index=168, g-index=370, 151,800+ citations).
\end{IEEEbiography}

\vfill

\end{document}